\documentclass[fleqn,usenatbib]{mnras}

\usepackage{newtxtext,newtxmath}

\usepackage[T1]{fontenc}

\DeclareRobustCommand{\VAN}[3]{#2}
\let\VANthebibliography\thebibliography
\def\thebibliography{\DeclareRobustCommand{\VAN}[3]{##3}\VANthebibliography}

\usepackage{graphicx}	
\usepackage{amsmath}	

\usepackage{microtype} 

\usepackage{etoolbox, siunitx}
\robustify\bfseries

\usepackage{booktabs} 

\usepackage[normalem]{ulem} 

\newcommand{\vsini}{\ensuremath{v\sin{i_{\star}}}}

\DeclareSIUnit\au{AU}
\DeclareSIUnit\Rsun{R_\odot}
\DeclareSIUnit\Rjup{R_\text{Jup}}
\DeclareSIUnit\Msun{M_\odot}
\DeclareSIUnit\Mjup{M_\text{Jup}}
\DeclareSIUnit\gyr{Gyr}
\DeclareSIUnit\ppt{ppt}
\DeclareSIUnit\ppm{ppm}

\title[TOI-1259B white dwarf spectroscopy]{Spectroscopy of TOI-1259B - an unpolluted white dwarf companion to an inflated warm Saturn}

\author[Fitzmaurice et al.]{%
        Evan Fitzmaurice$^{1}$$^{\href{https://orcid.org/0000-0003-0199-9699}{\includegraphics[scale=0.5]{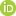}}}$,
        David V. Martin$^{1,2}$$^{\href{https://orcid.org/0000-0002-7595-6360}{\includegraphics[scale=0.5]{orcid.jpg}}}$,
        Romy Rodr\'iguez Mart\'inez$^{1}$$^{\href{https://orcid.org/0000-0003-1445-9923}{\includegraphics[scale=0.5]{orcid.jpg}}}$,\newauthor
        Patrick Vallely$^{1,3}$$^{\href{https://orcid.org/0000-0001-5661-7155}{\includegraphics[scale=0.5]{orcid.jpg}}}$, 
        Alexander P. Stephan$^{1,4}$$^{\href{https://orcid.org/0000-0001-8220-0548}{\includegraphics[scale=0.5]{orcid.jpg}}}$,
        Kiersten M. Boley$^{1,3}$$^{\href{https://orcid.org/0000-0001-8153-639X}{\includegraphics[scale=0.5]{orcid.jpg}}}$,
        Rick Pogge$^{1}$$^{\href{https://orcid.org/0000-0003-1435-3053}{\includegraphics[scale=0.5]{orcid.jpg}}}$,\newauthor
        Kareem El-Badry$^{5}$$^{\href{https://orcid.org/0000-0002-6871-1752}{\includegraphics[scale=0.5]{orcid.jpg}}}$,
        Vedad Kunovac$^{6}$$^{\href{https://orcid.org/0000-0001-9419-3736}{\includegraphics[scale=0.5]{orcid.jpg}}}$ \&
        Amaury H. M. J. Triaud$^{7}$$^{\href{https://orcid.org/0000-0002-5510-8751}{\includegraphics[scale=0.5]{orcid.jpg}}}$
\vspace{0.3cm}
\\
\vspace{0.3cm}
fitzmaurice.11@buckeyemail.osu.edu\\
$^{1}$Department of Astronomy, The Ohio State University, 4055 McPherson Laboratory, Columbus, OH 43210, USA\\
$^{2}$NASA Sagan Fellow\\
$^{3}$NSF Graduate Fellow\\
$^{4}$Center for Cosmology and AstroParticle Physics, The Ohio State
University, Columbus, OH 43210, USA \\
$^{5}$Center for Astrophysics | Harvard \& Smithsonian, 60 Garden St, Cambridge, MA 02138, USA \\
$^{6}$Lowell Observatory, 1400 W. Mars Hill Rd., Flagstaff, AZ 86001, USA\\
$^{7}$School of Physics and Astronomy, University of Birmingham, Edgbaston, Birmingham B15 2TT, UK\\
}

\vspace{-2cm}
\date{Accepted at MNRAS}

\pubyear{2022}

\begin{document}

\label{firstpage}
\pagerange{\pageref{firstpage}--\pageref{lastpage}}
\maketitle

\begin{abstract}

TOI-1259 consists of a transiting exoplanet orbiting a main sequence star, with a bound outer white dwarf companion. Less than a dozen systems with this architecture are known. We conduct follow-up spectroscopy on the white dwarf TOI-1259B using the Large Binocular Telescope (LBT) to better characterise it. We observe only strong hydrogen lines, making TOI-1259B a DA white dwarf. We see no evidence of heavy element pollution, which would have been evidence of planetary material around the white dwarf. Such pollution is seen in $\sim25-50\%$ of white dwarfs, but it is unknown if this rate is higher or lower in TOI-1259-like systems that contain a known planet. Our spectroscopy permits an improved white dwarf age measurement of $4.05^{+1.00}_{-0.42}$ Gyrs, which matches gyrochronology of the main sequence star. This is the first of an expanded sample of similar binaries that will allow us to calibrate these dating methods and provide a new perspective on planets in binaries.

\end{abstract}




\section{INTRODUCTION}\label{sec:introduction}

Roughly $25-50$ percent of white dwarfs (WD) have elements heavier than helium in their upper atmospheres \citep{Zuckerman2010,Koester2014}, such as silicon, magnesium, iron, carbon and oxygen.  This ``pollution'', is unexpected, as the high gravity of a white dwarf should cause these metals to gravitationally settle into the interior quickly (timescales of weeks to a few thousand years, \citealt{Paquette1986}). Unless our observations are unrealistically well-timed, it is argued that this pollution must be constantly replenished \citep{Aannestad1993,Koester2009}. Potential sources of replenishment include a disintegrating planet \citep{vanderburg2015,gansicke2019,Buchan2022}, a circumstellar disk \citep{farihi2016} or the bombardment of planetesimals/asteroids \citep{stephan2017,Petrovich2017}. Whilst the specific source of the pollution is unknown and likely varied, broadly-speaking it is believed to be  planetary \citep{Veras2021} and we even have direct evidence of WD accretion \citep{Cunningham2022}.

In this paper we conduct spectroscopy of the white dwarf TOI-1259B. The white dwarf was both discovered and determined to be in a stellar binary by \citet{elbadry2019}. It is separated from the main sequence K-dwarf TOI-1259A by 13.9", which corresponds to a projected separation of 1648 AU (Fig.~\ref{fig:summary}). It was later found by \citet{martin2021} that TOI-1259A hosts a transiting inflated warm Saturn ($a_{\rm p}/R_\star$=12.3, $M_{\rm p}=0.44M_{\rm Jup}$, $R_{\rm p}=1.02R_{\rm Jup}$)\footnote{Not to be confused with WD 1856+534, the only known case of a WD itself hosting a bonafide, intact transiting planet \citep{vanderburg2020}.}. Only about a dozen systems are known with a main sequence star, an orbiting exoplanet and a bound white dwarf companion. Only in a few of these cases does the planet transit (review in \citealt{martin2021}). To our knowledge only three of these WDs have received spectroscopy (WASP-98, \citealt{southworth2020}; HD 27442, \citealt{chauvin2006,mugrauer2007} and HD 107148 \citealt{mugrauer2016}) and no pollution has ever been detected (see Sect.~\ref{subsec:discussion_other_binaries}).

There are two motivations for this paper. First, we are looking for pollution in the WD. Since this would be a proxy for a planet, and the WD is bound to a known exoplanet, this is a novel means of probing planets around both stars of a wide binary. Only two binaries are known where both stars individually\footnote{As opposed to circumbinary planets, where the tight inner binary collectively hosts an outer planet (review in \citealt{martin2018}).} host a planet: WASP-94 \citep{neveuvanmalle2014} and XO-2 \citep{Damasso2015}. There have been many studies on the effect of stellar multiplicity on  planet occurrence rates \citep{Roell2012,martin2015,kraus2016,martin2018}, with planets being less likely in binaries tighter than $\approx200$ AU \citep{moe2019}. However, such studies only concern planets around one of the stars, not both. Our second motivation is to spectroscopically characterise the WD such that we may better measure its age through the white dwarf cooling timescale. This may be compared to other age-dating methods of gyrochronology and isochrone fitting.

Our paper is structured as follows. Sect.~\ref{sec:observations} details the observations using the MODS spectrograph at the LBT. In Sect.~\ref{sec:analysis_and_results} we characterise both the WD  and the exoplanet host star. The results are discussed in Sect.~\ref{sec:discussion} before concluding in Sect.~\ref{sec:conclusion}.

\section{Observations}\label{sec:observations}

\begin{figure*}
    \centering
    \includegraphics[width=0.99\textwidth]{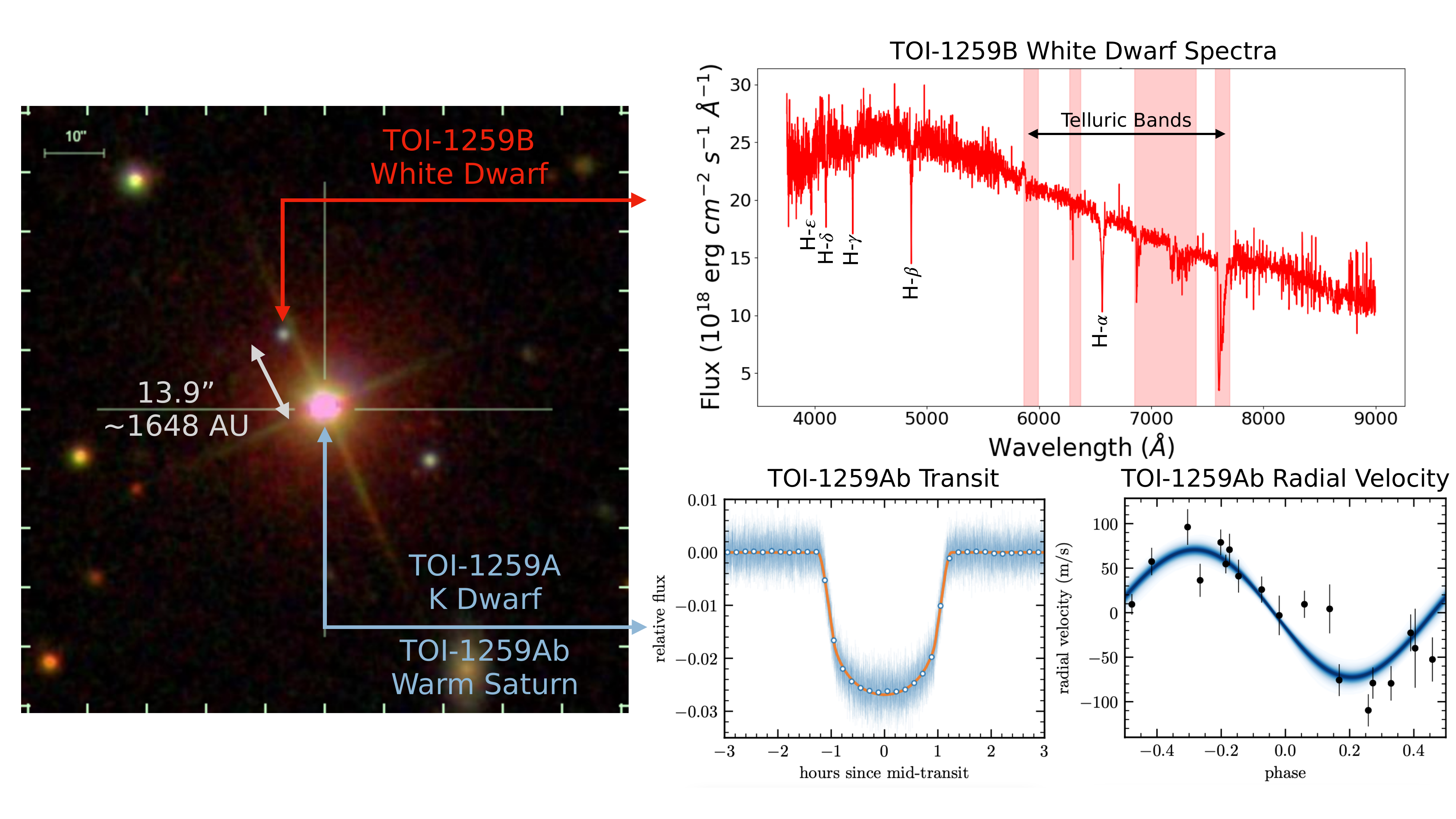}
    \caption{The TOI-1259 system, with a transiting warm Saturn and a bound WD companion. {\bf Left:} SDSS image showing the stellar binary. {\bf Top Right:} One-dimensional reduced spectrum of the DA white dwarf TOI-1259B. Outlier data was removed by visual inspection and manual cuts. Light red vertical bands are telluric absorption regions. Hydrogen features in the Balmer series are labeled. {\bf Bottom Right, Left Panel:} Phase folded and detrended TESS light curve of the transiting warm Saturn from \citet{martin2021}. {\bf Bottom Right, Right Panel:} RV data of the warm Saturn from SOPHIE in black and 99th percentile models in the blue region from \citet{martin2021}.}
    \label{fig:summary}
\end{figure*}

The TOI-1259 system contains two stars: a 12.08 Vmag K-dwarf and a 19.23 Vmag white dwarf, separated by 13.9" (SDSS image in Fig.~\ref{fig:summary}). We obtained spectra with both using the Multi-Object Double Spectrographs \citep[MODS;][]{PoggeMODS} mounted on the twin 8.4m Large Binocular Telescope.
All observations were performed in longslit mode using a 0.6{\arcsec} slit, oriented such that only one of the two stars was inside the slit for a given observation. MODS uses a dichroic that splits the light into separately optimized red and blue channels at 
$\sim5650$~{\AA}. For these observations we used the G400L reflection grating on the blue channel and the 
G670L reflection grating on the red channel. The MODS blue channels cover a wavelength range of 
$\sim3200-5650$~{\AA} with a resolution of $R\sim1850$ in the dual grating setup, while the MODS red
channels cover  $\sim5650-9800$~{\AA} with a resolution of $R\sim2300$. The spectra were reduced using the \textsc{modsccdred} \citep[][]{modsCCDRed}\footnote{https://github.com/rwpogge/modsCCDRed}\textsc{python} package for basic 2d CCD reductions, and the \textsc{modsidl} pipeline\citep[][]{modsIDL}\footnote{https://github.com/rwpogge/modsIDL} to extract 1d spectra and apply wavelength and flux calibrations. Obserations were also taken of the exoplanet host star TOI-1259A. two sequences of five 120-second exposures. Ultimately, these were not used in any analysis.


For the host star TOI-1259A we use existing SOPHIE spectra taken using the 1.93 m telescope at Observatoire de Haute Provence. Whilst it has significantly smaller (1.92m) aperture than the LBT, the higher resolution of SOPHIE ($R\sim$40,000 in High Efficiency mode) is needed to characterise TOI-1259A's metallicity. The data are publicly available \footnote{The ``objname'' in the SOPHIE database is GSC4591-0930. All spectra can be found here: \url{http://atlas.obs-hp.fr/sophie/sophie.cgi?n=sophies&a=htab&ob=ra,seq&c=o&o=GSC4591-0930}} and their acquisition is described in \citet{martin2021}.

\section{Analysis and results}\label{sec:analysis_and_results}

\subsection{Exoplanet host star TOI-1259A parameters}\label{subsec:analysis_host}

We determine the stellar parameters of the host star using the publicly-available EXOFASTv2 exoplanet suite \citep{eastman2013,Eastman2019}, which fits exoplanet data but has the capability of modeling  stellar-only properties. We perform a spectral energy distribution (SED) fit of the star with the \citet{Kurucz1992} atmosphere models and using broadband photometry; specifically, we use the Gaia $G$, $G_{BP}$ and $G_{RP}$ magnitudes, $JHK$ magnitudes from the 2MASS catalog \citep{Cutri2014}, and the W1-W4 magnitudes from WISE \citep{Zacharias2004}. To independently derive an age of the system, we combine the SED fit with three different stellar evolutionary models within EXOFASTv2. We first derive an age using the MESA Isochrones and Stellar Tracks (MIST;  \citealt{choi2016,dotter2016}) isochrones, which assume a non-rotating star. We also perform fits using the Yonsei-Yale (YY) evolutionary tracks \citep{Yi2001} and the PARSEC models \citep{Bressan1993}, which are similar to the MIST models.

For every fit, we impose initial guesses for the stellar mass $M_{\star}$, radius $R_{\star}$ and $T_{\rm eff}$ which are taken from the TESS Input Catalog (TIC-8; \citealt{stassun2019}). We adopt a prior on the visual extinction $A_{V}$ from the Schlegel dust maps \citep{schlegel1998} and a parallax from the Gaia DR3 \citep{gaia2020}. Based on our isochrone fitting we obtain approximate stellar ages of 5 Gyr, with uncertainties on the order of a few Gyr. These are weak constraints in comparison to gyrochronology and WD cooling.

We determine the [Fe/H] metallicity of the host star from a high-resolution ($R\sim$39,000) optical spectrum from the SOPHIE spectrograph, which spans the wavelength range $3870-6944$ nm. The spectrum we use has a signal-to-noise ratio of $\sim$20, which is the highest in our observations. We use the open source framework for spectral analysis, {\tt iSpec} \citep{blanco-cuaresma14, blanco-cuaresma19}. {\tt iSpec} calculates stellar atmospheric parameters as well as individual abundances using the synthetic spectral fitting method and the equivalent width method. For this target, we chose the synthetic spectral fitting method, as recommended by \citet{blanco-cuaresma14}. Within the {\tt iSpec} framework, we employ the radiative transfer code SPECTRUM \citep{Gray1994}, the solar abundances from \citet{Grevesse2007}, the MARCS.GES atmosphere models \citep{Gustafsson2008}, and the atomic line list from the Gaia-ESO Survey (GES). The stellar parameters required for the abundance determinations, namely, the stellar effective temperature and surface gravity, were taken from our EXOFASTv2 analysis. These were $T_{\rm eff} = 4747^{+72}_{-74}$ K, and $\log{g_*} = 4.6 \pm 0.025$. We fix the limb darkening coefficient to 0.6, and we fix the stellar rotation, \vsini, to 2 km/s (which is a typical value for sun-like stars) to avoid possible degeneracies between the macroturbulence and rotation, following the recommendations of \citet{blanco-cuaresma14}. We determine a final metallicity of [Fe/H] = $-0.1 \pm 0.04$ dex.

\subsection{White dwarf companion TOI-1259B parameters}\label{subsec:subsec:analysis_wd}


The reduced one-dimensional spectrum of the white dwarf is shown in the top right panel of Fig.~\ref{fig:summary}. Outlier data were removed by visual inspection. The light red bands indicate regions where telluric features are expected. Hydrogen features in the Balmer series that are present are labeled (H-$\alpha$, -$\beta$, -$\gamma$, -$\delta$, -$\epsilon$). The spectrum lacks any features of helium or metals. Because of the presence of the hydrogen features and the absence of other lines, we classify this white dwarf as a DA white dwarf. 

We use the open source package \texttt{wdtools} \citep{Chandra2020}\footnote{\url{https://wdtools.readthedocs.io/}}, which is designed to fit DA white dwarf spectra and derive the effective temperature, T$_{\rm eff}$, and surface gravity, log\textit{g}.  This software interfaces with theoretical spectra computed by \cite{Tremblay2009,Koester2010}. We use the Generative Fitting Pipeline (GFP) method. To build this method \citet{Chandra2020} use private state-of-the-art atmospheric models to create initial synthetic spectra and fit them to SDSS spectra. For computational efficiency they use a neural network adapted for white dwarf spectra to generate more synthetic spectra as a part of the fitting process. They reserve 1 percent of their SDSS spectra for validation of the neural network model and obtain a relative error of less than $0.01$ in normalized flux units across all pixels. 

With the ability to generate unlimited high-quality synthetic spectra, they utilize an MCMC algorithm to sample the posterior and select the highest likelihood (lowest $\chi_r^2$) fit for the labels of effective temperature and log\textit{g}. The MCMC is able to take in a prior for the effective temperature in order to inform the fit.   

\begin{figure}
    \centering
    \includegraphics[width=0.47\textwidth]{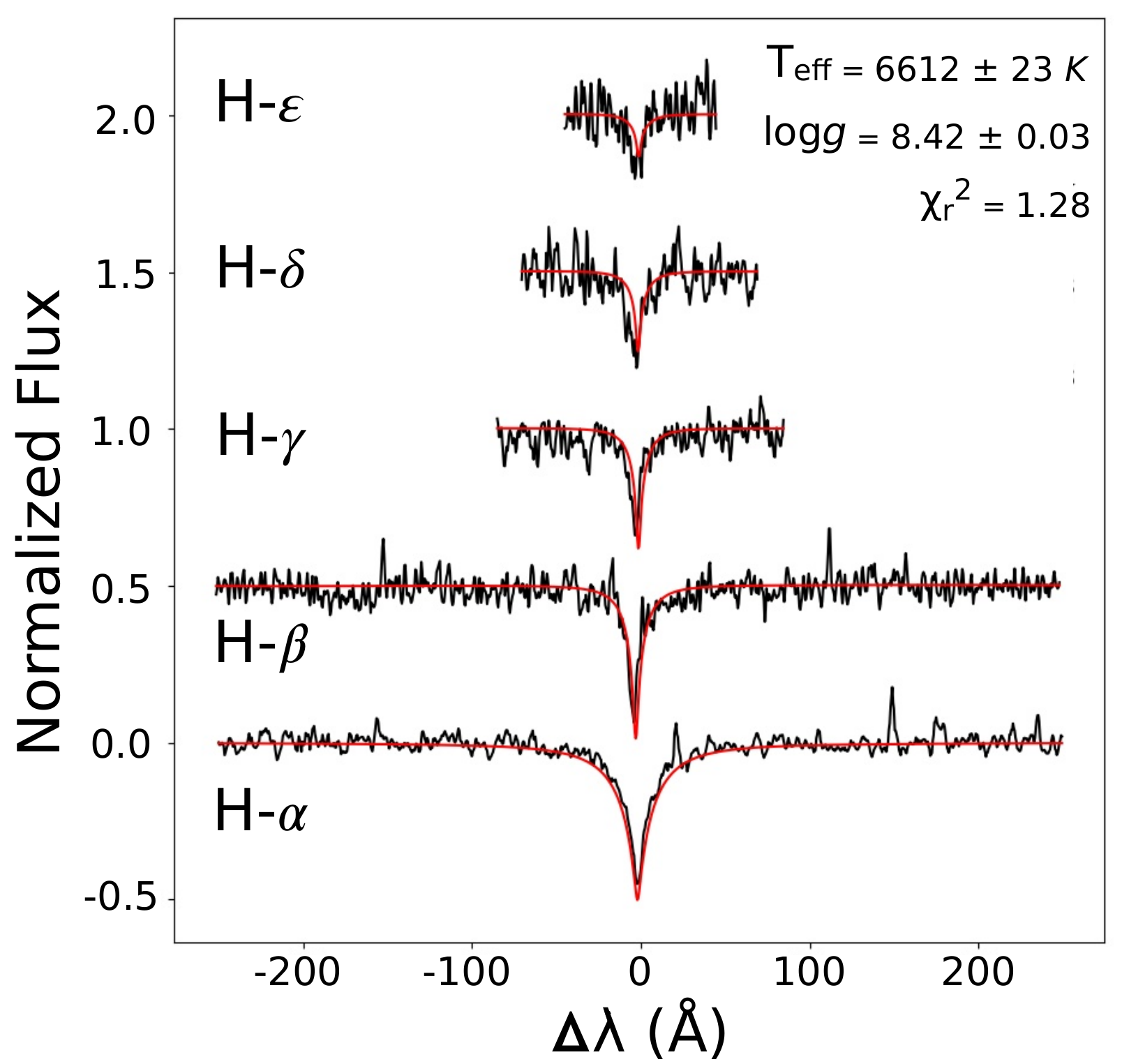}
    \caption{Spectroscopy of the DA white dwarf TOI-1259B. We determine $T_{\rm eff}$ and log\textit{g} from fitting the H-$\alpha$, $-\beta$, $-\gamma$, $-\delta$, and $-\epsilon$ lines using \texttt{wdtools} \citep{Chandra2020}.}
    \label{fig:lines_fit}
\end{figure}

We first scale the flux and flux errors by a factor of $10^{18}$ so that the program can accurately fit the continuum and normalize the spectrum. We inform the pipeline of the instrument spectral dispersion in angstroms. For MODS, this value is different for the red and blue channels that are used, being $0.8$ and $0.5$ respectively. We use an intermediate value of $0.65$ since the data is combined in reduction. We inform the pipeline of the prior effective temperature of $T_{\rm eff} = 6300^{+80}_{-70}$ K that is derived from photometry in \citet{martin2021}. There is another prior from \citet{mugrauer2020} of $T_{\rm eff} = 6473^{+672}_{-419}$ K that we test but do not use as there is not a significant change in the produced values or fit. The pipeline allows for selection of specific hydrogen Balmer features to be used in the fit, so we elect to use the H-$\alpha$, $-\beta$, $-\gamma$, $-\delta$, and $-\epsilon$ features since they are all present in the spectrum. The produced spectroscopic fit and labels are shown in Fig. \ref{fig:lines_fit}. We derive an effective temperature of $T_{\rm eff} = 6612 \pm 23 $ K, and a log$g = 8.42 \pm 0.03$.

For comparison, we re-create the SED fit of the WD by following \citet{martin2021} but with two improvements. First, the biggest improvement is that we are now certain it is a DA, whereas previously there was a systematic uncertainty in the ages due to our ignorance of the WD spectral type. Second, we provide a metallicity prior of [Fe/H] = $-0.1 \pm 0.04$ dex based on spectroscopic analysis of the main sequence TOI-1259A. This is an improvement on the rough [Fe/H] = $-0.2 \pm 0.3$ dex used in \citet{martin2021}, taken from the Tess Input Catalog (TIC). Like in \citet{martin2021} we use two different WD Initial Final Mass Relations (IFMR): \citet{williams2009} and \citet{elbadry2018IFMR}. 

We derive an SED temperature of $T_{\rm eff}=6330\pm70$ K, which differs slightly from the spectroscopic temperature of $T_{\rm eff} = 6612 \pm 23 $ K. Typically, in the era of Gaia-derived distances the SED result should be considered more trustworthy. To first approximation, the SED just relies on the Stefan-Boltzmann law, whereas the interpretation of the WD spectrum relies on complicated non-LTE radiative transfer calculations \citep{Beaulieu2019,Bergeron2019}. Our age constraints are slightly tighter than in \citet{martin2021}, with  $3.74^{+0.50}_{-0.22}$ using the \citet{williams2009} and $4.05^{+1.00}_{-0.42}$ Gyr using \citet{elbadry2018IFMR}.  The mass of the WD in these fits is $0.57\pm 0.02\,M_{\odot}$.

 The constraints we derive on the WD's age represent formal fitting uncertainties, and do not account for systematics such as uncertainties in the WD cooling models, uncertainties in the IFMR, or problems with the observational data. The fact that our inferred spectroscopic and photometric temperatures for the WD are formally inconsistent suggests that the errors on both quantities could be underestestimated. We performed two fitting experiments to assess the effects of systematics and underestimated uncertainties on our age constraints. 

First, we tried inflating the SDSS photometric uncertainties until our photometric and spectroscopic temperature constraints were consistent. We found that this was achieved when we inflated the uncertainty in all bands to 0.08 mag (where the reported values are $\sim$0.02 mag). Re-running BASE-9 with these inflated uncertainties and the \citet{elbadry2018IFMR} IFMR, we found a photometric temperature of $6459 \pm 140$ K, which is consistent within 1 sigma with the spectroscopic value. In this case, the inferred total age changes from $4.05_{-0.42}^{+1.00}$ Gyr, to $3.62_{-0.30}^{+1.52}$ Gyr. As expected, this value is consistent with what we inferred from the reported uncertainties but has a somewhat larger errorbar.

Next, we removed the photometry from the fit and instead put a tight prior on the WD temperature and surface gravity from our spectroscopic fit. In this case, we obtained a total age of $3.01_{-0.09}^{+0.32}$ Gyr, which is inconsistent at the $\sim2$ sigma level with our nominal uncertainty. This is not surprising, since the photometric and spectroscopic effective temperatures differ at the $\sim 3$ sigma level.

In summary, all the WD age fitting approaches we explored place it an an age between about 2.9 and 5 Gyr. The differences between ages inferred under different assumptions are comparable to the formal fitting uncertainty under our fiducial assumptions, and thus do contribute to to the total realistic uncertainty budget. This is not unexpected for a relatively low-mass WD such as TOI-1259B, for a small change in inferred age can translate to a significant change in both cooling and pre-WD age \citep[e.g.][]{Heintz2022}.


\section{Discussion}\label{sec:discussion}

\subsection{Comparative age measurements}

\renewcommand{\arraystretch}{1.5}
\begin{table}

    \centering
    \caption{Age measurements for the TOI-1259 based on different methods.}  
    \footnotesize
    \begin{tabular*}{\columnwidth}{@{\extracolsep{\fill}}
        l
        l
        l
        }
        \toprule
        \toprule
        Method & Age  & Error \\
         & (Gyr)  & (Gyr) \\
        \midrule

        1. Isochrone (MIST) & 4.9 &  $^{+5.5}_{-3.5}$ \\
        2. Isochrone (Yonsei-Yale) & 5.3 & $^{+4.8}_{-3.7}$ \\
        3. Isochrone (PARSEC) & 5.2 & $^{+5.2}_{-3.7}$ \\
        4. WD Cooling (\citealt{williams2009} IFMR) & 3.74 & $^{+0.50}_{-0.22}$ \\
        5. WD Cooling (\citealt{elbadry2018IFMR} IFMR) & 4.05 & $^{+1.00}_{-0.42}$ \\
        6. Gyrochronology & 4.8 & $^{+0.7}_{-0.8}$ \\
        
        \bottomrule
    \end{tabular*}
    \label{tab:summary}
\end{table}

Table~\ref{tab:summary} shows the age constraints from the three isochrones tested (Sect.~\ref{subsec:analysis_host}), the WD cooling age (Sect.~\ref{subsec:subsec:analysis_wd}) and gyrochronology age derived in \citet{martin2021}. The gyrochronological age, which we do not recalculate here, is based on a Systematics-Insensitive Periodogram from \citet{Hedges_2020} where astrophysical periodicities in the light curve are isolated with respect to background variations due to the spacecraft. This rotation rate is converted to an age following the methodology of \citet{angus2020}.

For the WD ages there are two components: the time spent as a white dwarf (the cooling time) and the time spent before becoming a white dwarf (essentially its main sequence age). The cooling age is very well constrained: $1.89^{+0.07}_{-0.06}$ Gyr. The pre-white dwarf age is much less certain because the time a star spends on the main sequence is a sensitive function of stellar mass, and we have to estimate the main sequence progenitor's mass based on the current white dwarf mass. This process also suffers from a systematic uncertainty related to the choice of the IFMR (\citealt{williams2009} or \citealt{elbadry2018IFMR}), which is why we provide both ages in Table~\ref{tab:summary}. 

Our WD age measurements are improved in comparison with \cite{martin2021} in two ways. First, the by confirming TOI-1259B as a DA white dwarf we have removed a large source of systematic uncertainty of the WD spectral type. This uncertainty propagates into the stellar parameters and ultimately the age. The second improvement comes from better constraining the metallicity of the white dwarf progenitor, since main sequence age is also affected by metallicity. This comes through spectral characterisation of the main sequence TOI-1259A, which we presume has the same metallicity. Overall, we have a  $\sim20\%$ tighter age constraint than \cite{martin2021}.

For completeness, we include in Table~\ref{tab:summary} the stellar ages derived from our isochrone fits to the primary star TOI-1259A (MIST, Yonsei-Yale and PARSEC). However, given the low mass ($M_{\rm A}=0.744M_\odot$) and hence long expected life-time for this K-dwarf, the isochrones provide very little constraint on its age, as reflected in the large errorbars. The gyrochronology and WD cooling ages are much more constraining, with $\delta{\rm t_{\rm Age}}\approx 20\%$, and it is re-assuring to see that these two methods match. An expanded sample of WD + main sequence binaries will provide a more thorough comparison and calibration of these two age dating methods.

\subsection{Lack of pollution in context of TOI-1259's dynamical history}\label{subsec:discussion_nopollution}

\citet{Zuckerman2014} determined that the rate of WD pollution was the same in single WDs as in  WDs in wide ($>1000$ AU) binaries. For tighter binaries the WD pollution rate was decreased. This may indicate that in these tighter binaries planet formation is suppressed due to binary interactions. This matches exoplanets in main sequence binaries both in terms of  observations \citep{kraus2016,Su2021} and theory \citep{Kley2008,Marzari2019}. However, in the  \citet{Zuckerman2014} sample the main sequence star in the binary was not known to host a planet. It is unknown what the pollution rate is in TOI-1259-like architectures.

Whilst the WD spectrum produced from our observations shows no signs of such pollution in TOI-1259B, the absence of such lines does not conclusively rule out that this WD has not ever had planets or has not ever been polluted by planetary material in the past. Specifically, given the age and architecture of the TOI-1259 system, a variety of scenarios may explain the current absence of metal lines, while still allowing for frequent pollution events. 

For example, the presence of the stellar companion may have driven most planetary material into the white dwarf progenitor during the main sequence and red giant phases via the Kozai-Lidov mechanism \citep{kozai1962,lidov1962,naoz2016,Martin2016}, with only a small amount of material available for accretion for the WD \citep[a scenario discussed in][]{stephan2017}. These pollution events may also only occur sporadically, with accreted material quickly settling into the deeper WD layers, avoiding detection.  For our WD we expect a settling timescale of $\approx5000-10,000$ years, which is much shorter than the cooling timescale \citep{Paquette1986,Bedard2020}.  As the Kozai-Lidov mechanism is only effective on bodies in specific parts of the eccentricity-inclination parameter space, there may also still be significant amounts of planetary material left on orbits that do not allow them to come close to the WD. 

In order to determine if the presence of a stellar companion aids or hinders to pollute WDs with planetary material, a larger sample of WDs in binaries will have to be observed to allow comparisons with the expected value of $25-50$ percent from single WD observations.

\subsection{Comparison with other white dwarfs in binaries}\label{subsec:discussion_other_binaries}
Previous works have performed similar analysis on the spectroscopy and photometry of white dwarf companions to exoplanet hosting stars. \citet{southworth2020} analyzed the spectra of the white dwarf WASP-98B and found a featureless spectrum which prohibited a precise age estimation. \citet{mugrauer2016} analyzed the spectra of HD 107148B and determined a spectral type DA. The effective temperature was found to be between 5900 and 6400 K, and they deduced a total age of the system to be $6.0 \pm 4.8$ Gyr. \citet{mugrauer2007} and \citet{chauvin2006} both analyzed the white dwarf companion HD 27442B. This white dwarf is classified as a DA WD with an effective temperature $\sim$14400 K and a cooling age of $\sim$220 Myr, but a total age is not computed. \citet{vanderburg2015a} did photometric analysis of HIP 116454B using SDSS \textit{ugriz} data and determined an effective temperature of $7500 \pm 200 K$ and cooling age of $\sim$1.3 Gyr. TOI-1259 stands out in this sample in that not only do we have a WD cooling age but we also have a gyrochronology age.

Stellar age calibration for WD-MS binaries has been applied for many different scenarios. \citet{Catalan2008b, Cummings2018,williams2009} use white dwarfs in binaries to refine initial-final mass relations by using spectroscopic and photometric measurements of the white dwarf and its companion to determine the white dwarf cooling age, as well as the progenitor mass as determined from the estimated age of the main sequence companion. \citet{Fouesneau2019} and \citet{Qiu2021} pair spectroscopic and distance measurements to determine ages of field stars with white dwarf companions that are otherwise difficult to determine. \citet{Rebassa-Mansergas2021} uses this method to refine age measurements of stars in the solar neighborhood to determine the age-metallicity relation. \citet{Garces2011} tracks the decrease in high-energy emission of low mass stars by using white dwarf companions to calibrate the ages. This method of age calibration using white dwarf companions is broadly applicable and widely trusted, and serves as a reliable mechanism for our purpose. 

\section{Conclusion}\label{sec:conclusion}

We use the MODS spectrograph on the LBT to observe the white dwarf TOI-1259B, which is a bound companion to the inflated warm Saturn TOI-1259b. We observe solely strong Hydrogen lines, characterising it as a DA WD. No heavy element pollution is seen in the WD's atmosphere, which would have been evidence of planets (or at least planetary material) around both stars in a wide binary, a phenomena for which there exists only minimal observational constraints. Whilst WD pollution is seen in roughly $25-50\%$ of single WDs \citep{Zuckerman2010,Koester2014}, and similarly in $>1000$ binaries \citep{Zuckerman2014}, we have no constraints on the rate of WD pollution in TOI-1259-like architectures where the main sequence star hosts a planet.

The TOI-1259 system remains interesting for future follow-up. As an inflated warm Saturn ($a_{\rm p}/R_\star=12.3$, $M_{\rm p}=0.44 M_{\rm Jup}$, $R_{\rm p}=1.02 R_{\rm Jup}$) with very deep $2.7\%$ transits around a K-dwarf it is an ideal target for atmospheric characterisation \citep{kempton2018}. Unlike most exoplanets, the WD companion provides a reliable and precise ($\sim20\%$) age measurement. Therefore, any future observations that would speak to a dynamical past, such as companion planets, spin-orbit obliquities or residual eccentricity, could be better interpreted. 

\section*{Acknowledgements}

This paper made use of data obtained with the Large Binocular Telescope (LBT). The LBT is an 
international collaboration among institutions in the United States, Italy and Germany. LBT Corporation 
partners are: The University of Arizona on behalf of the Arizona Board of Regents; Istituto Nazionale di 
Astrofisica, Italy; LBT Beteiligungsgesellschaft, Germany, representing the Max-Planck Society, The 
Leibniz Institute for Astrophysics Potsdam, and Heidelberg University; The Ohio State University, 
representing OSU, University of Notre Dame, University of Minnesota and University of Virginia.

LBT MODS data in this paper used the modsIDL spectral data reduction pipeline and the modsCCDRed data
reduction code developed in part with funds provided by NSF Grants AST-9987045 and AST-1108693, and a
gift from David G. Price through the Price Fellowship in Astronomical Instrumentation. 

This project used funding from the NASA TESS Guest Investigator Program (G04157, PI Martin). Support for DVM's work was provided by NASA through the
NASA Hubble Fellowship grant HF2-51464 awarded by the Space Telescope Science Institute, which is operated by the Association of Universities for Research in Astronomy, Inc., for NASA, under contract NAS5-26555. V.K. acknowledges support from NSF award AST2009501.


\section*{Data availability}
The SOPHIE and MODS spectroscopy will be made publicly available. Any data fits can be provided on request to the authors.

\bibliographystyle{mnras}
\bibliography{references}

\appendix

\bsp	
\label{lastpage}
\end{document}